\documentstyle[twoside]{article}

\catcode`\@=11
\long\def\@makefntext#1{
\protect\noindent \hbox to 3.2pt {\hskip-.9pt  
$^{{\eightrm\@thefnmark}}$\hfil}#1\hfill}		

\def\@makefnmark{\hbox to 0pt{$^{\@thefnmark}$\hss}}	
	
\def\ps@myheadings{\let\@mkboth\@gobbletwo
\def\@oddhead{\hbox{}
\rightmark\hfil\eightrm\thepage}   
\def\@oddfoot{}\def\@evenhead{\eightrm\thepage\hfil
\leftmark\hbox{}}\def\@evenfoot{}
\def\sectionmark##1{}\def\subsectionmark##1{}}

\oddsidemargin=\evensidemargin
\newcommand{\textlineskip}{\baselineskip=13pt}
\newcommand{\smalllineskip}{\baselineskip=10pt}

\def\eightcirc{
\begin{picture}(0,0)
\put(4.4,1.8){\circle{6.5}}
\end{picture}}
\def\eightcopyright{\eightcirc\kern2.7pt\hbox{\eightrm c}}

\def\abstracts#1#2#3{{
    \centering{\begin{minipage}{4.5in}\baselineskip=10pt\footnotesize
    \parindent=0pt #1\par 
    \parindent=15pt #2\par
    \parindent=15pt #3
    \end{minipage}}\par}} 

\renewenvironment{thebibliography}[1]
    {\frenchspacing
    \ninerm\baselineskip=11pt
    \begin{list}{\arabic{enumi}.}
    {\usecounter{enumi}\setlength{\parsep}{0pt}
    \setlength{\leftmargin 12.7pt}{\rightmargin 0pt} 
    \setlength{\itemsep}{0pt} \settowidth
    {\labelwidth}{#1.}\sloppy}}{\end{list}}

\newcounter{itemlistc}

\newcommand{\tcaption}[1]{
        \refstepcounter{table}
        \setbox\@tempboxa = \hbox{\footnotesize Table~\thetable. #1}
        \ifdim \wd\@tempboxa > 5in
           {\begin{center}
     \parbox{6.3in}{\footnotesize\smalllineskip \hspace{-0.3in}
           Table~\thetable. #1}
           \end{center}}
        \else
             {\begin{center}
             {\footnotesize Table~\thetable. #1}
              \end{center}}
        \fi}

\def\@citex[#1]#2{\if@filesw\immediate\write\@auxout
	{\string\citation{#2}}\fi
\def\@citea{}\@cite{\@for\@citeb:=#2\do
	{\@citea\def\@citea{,}\@ifundefined
	{b@\@citeb}{{\bf ?}\@warning
	{Citation `\@citeb' on page \thepage \space undefined}}
	{\csname b@\@citeb\endcsname}}}{#1}}

\newif\if@cghi
\def\cite{\@cghitrue\@ifnextchar [{\@tempswatrue
	\@citex}{\@tempswafalse\@citex[]}}
\def\citelow{\@cghifalse\@ifnextchar [{\@tempswatrue
	\@citex}{\@tempswafalse\@citex[]}}
\def\@cite#1#2{{$\null^{\hspace{-0.03in}#1}$\if@tempswa\typeout
	{IJCGA warning: optional citation argument 
	ignored: `#2'} \fi}}

\def\pmb#1{\setbox0=\hbox{#1}
	\kern-.025em\copy0\kern-\wd0
	\kern.05em\copy0\kern-\wd0
	\kern-.025em\raise.0433em\box0}

\def\fpage#1{\begingroup
\voffset=.3in
\thispagestyle{empty}\begin{table}[b]\centerline{\footnotesize #1}
	\end{table}\endgroup}

\def\runninghead#1#2{\pagestyle{myheadings}
\markboth{{\protect\footnotesize\it{\quad #1}}\hfill}
{\hfill{\protect\footnotesize\it{#2\quad}}}}
\font\ninerm=cmr9

\font\eightrm=cmr8

\newcommand{\dzero}     {D\O}
\newcommand{\rar}       {\rightarrow}
\newcommand{\rargap}    {\mbox{ $\rightarrow$ }}
\newcommand{\ttbar}     {\mbox{$t\bar{t}$}}
\newcommand{\bbbar}     {\mbox{$b\bar{b}$}}
\newcommand{\ppbar}     {\mbox{$p\bar{p}$}}
\newcommand{\Wbb}       {\mbox{$Wb\bar{b}$}}
\newcommand{\Wcc}       {\mbox{$Wc\bar{c}$}}
\newcommand{\Wss}       {\mbox{$Ws\bar{s}$}}
\newcommand{\Vtb}       {\mbox{$V_{tb}$}}

\begin{document}

\runninghead{Search for Electroweak Production of Single Top Quarks at
{\dzero}} {Search for Electroweak Production of Single Top Quarks at
{\dzero}}

\normalsize\textlineskip
\thispagestyle{empty}
\setcounter{page}{1}


\begin{flushright}
{\dzero} Note 3816 ~~ UCR/{\dzero}/01-01

\vspace{-0.015in}
Fermilab-Conf-01/004-E

\vspace{-0.015in}
October 2000
\end{flushright}

\vspace*{0.4truein}

\fpage{1}
\centerline{\bf SEARCH FOR ELECTROWEAK PRODUCTION}
\vspace*{0.035truein}
\centerline{\bf OF SINGLE TOP QUARKS AT {\dzero}}
\vspace*{0.32truein}
\centerline{\footnotesize A.P. HEINSON}
\vspace*{0.015truein}
\centerline{\footnotesize\it {Department of Physics, University of California}}
\baselineskip=10pt
\centerline{\footnotesize\it Riverside, CA 92521-0413, USA}
\vspace*{10pt}
\centerline{\footnotesize FOR THE {\dzero} COLLABORATION}

\vspace*{0.21truein}
\abstracts{This paper discusses a search for electroweak production
of single top quarks in the electron+jets and muon+jets decay
channels. The measurements use $\approx90$~pb$^{-1}$ of data from
Run~1 of the Fermilab Tevatron collider, collected at 1.8~TeV with the
{\dzero} detector. We use events that include a tagging muon, implying
the presence of a $b$ jet, to set an upper limit at the $95\%$
confidence level on the cross section for the $s$-channel process
${\ppbar}{\rargap}tb+X$ of 39~pb. The upper limit for the $t$-channel
process ${\ppbar}{\rargap}tqb+X$ is 58~pb.}{}{}

\textlineskip
\vspace*{12pt}

\noindent The {\dzero} collaboration recently completed a search for
single top quarks produced in association with a bottom quark or with
a light quark and a low-$p_T$ $b$~quark.~\cite{dzeropaper} The CDF
collaboration has reported similar measurements.~\cite{cdfpaper} These
analyses search for two independent modes that produce single top
quarks: the $s$-channel process $q^{\prime}\bar{q}{\rar}tb$ with a
predicted cross section
{\mbox{$\sigma=0.73\pm0.04$~pb};~\cite{willenbrock-s}} and the
$t$-channel process $q^{\prime}g{\rar}tqb$ with
{\mbox{$\sigma=1.70\pm0.19$~pb.~\cite{willenbrock-t}} We use the
notation ``$tb$'' to refer to both $t\bar{b}$ and the charge-conjugate
process $\bar{t}b$, and ``$tqb$'' for both $tq\bar{b}$
and~$\bar{t}\bar{q}b$.  Events are identified by the presence of one
isolated electron or muon and missing transverse momentum assumed to
be from the decay of a $W$~boson to a lepton and neutrino. The events
must also contain two to four jets, with one or more having an
associated muon to tag it as a possible $b$~jet.

The principal difficulty in undertaking a search for single top quarks
is that there are many other processes with similar topologies, but
far higher cross sections. The single top quark processes represent a
combined cross section of 2.4~pb, which corresponds to about one in
$10^{10}$ interactions. After selecting interesting events with
suitable triggers, the percentage of signal in $e$+jets or $\mu$+jets
decays is increased to $0.002\%$. Application of particle
identification criteria and judicious selections to reject the
backgrounds increases the combined percentage to $4\%$. This is the
best that can be done using simple event-selection techniques and the
modest amount of data currently available to {\dzero}.

\vspace{0.2in}
\footnoterule
\small
\noindent Presented at the Meeting of the Division of Particles
and Fields of the American Physical Society, The Ohio State
University, Columbus, Ohio, 9th--12th August 2000.
\normalsize

\clearpage
\textheight=7.8truein
\topmargin=0truein

Improving the signal selection and background rejection in the search
for single top quark events is important, since observation of this
complementary production mode would add significantly to the knowledge
of the top quark obtained from studies of {\ttbar}
pairs.~\cite{ttreview} For instance, the cross section for the
electroweak production of top quarks determines the magnitude of the
CKM matrix element {\Vtb}.~\cite{measVtb} Two avenues are being
pursued to enhance the search for single top quark production. We
expect to increase the amount of data by a factor of $\approx 20$ in
Tevatron Run~2. The quality of the information associated with each
event will be improved significantly, owing to the addition of the
Silicon Microstrip Tracker for reconstructing secondary vertices in
$b$~jets, and from improvements in electron and muon identification
systems. Before these data become available, we can also enhance
signal selection techniques with the use of neural networks, which
will enable us to extend the search into events without a tagging
muon.

A detailed understanding of the backgrounds in the single top quark
channels can also be used to increase signal acceptance and background
rejection, and efforts are being focused on areas of the analysis that
can lead to significant improvements. In this paper, we discuss the
backgrounds after imposition of sequentially tighter signal
selections, and note where improvements of the analysis in Run~2
appear likely. Since the final state particles in single top quark
events are the same as for associated Higgs boson production
$q^{\prime}\bar{q}{\rar}WH$ with $H{\rar}{\bbbar}$, understanding the
backgrounds to single top quark production will also help in the
investigation of this important process.

\begin{table}[!h!tbp]
\tcaption{Percentages of signal and background events predicted after
the imposition of each set of criteria.}
\label{table1}
\centerline{\footnotesize\smalllineskip
\begin{tabular}{lcccccccccc} \\
\hline
             &  $tb$   & $tqb$  & {\ttbar}& {\Wbb} & {\Wcc}& $Wjj$&  $WW$  &  $WZ$  & QCD &Other\\
\hline
\underline{Electron Chan.} &&&&&&&&&& \\
~~Trigger    & 0.00050 & 0.0012 &  0.0083 & 0.0038 & 0.017 &  5.7 & 0.0094 & 0.0013 & 94  & --- \\
~~Baseline   & 0.20    & 0.34   &  5.3    & 0.65   & 0.72  &  3.1 & 0.43   & 0.13   & 89  & --- \\
~~Loose      & 0.75    & 1.3    & 11      & 3.8    & 4.4   & 17   & 2.2    & 0.73   & 59  & --- \\
~~Tight      & 1.4     & 2.1    &  8.7    & 6.3    & 6.8   & 25   & 3.4    & 1.2    & 45  & ---
\vspace{0.02 in} \\
\underline{Muon Channel}   &&&&&&&&&& \\
~~Trigger    & 0.00093 & 0.0022 &  0.015  & 0.0055 & 0.023 &  7.4 & 0.020  & 0.0029 &  ?  & 93  \\
~~Baseline   & 0.15    & 0.27   &  4.0    & 1.3    & 1.3   &  3.9 & 1.2    & 0.25   & 5.7 & 82  \\
~~Loose      & 1.0     & 2.0    & 15      & 3.8    & 3.8   &  8.6 & 3.6    & 0.60   & 8.0 & 54  \\
~~Tight      & 1.6     & 2.6    &  9.0    & 5.3    & 5.3   &  4.8 & 5.7    & 1.3    & 8.0 & 56  \\
\hline
\end{tabular}}
\end{table}

Table~\ref{table1} shows the percentages of events for the two signals
and eight separate backgrounds that remain in the {\dzero} Run~1 data
after four stages of event selection: triggering ({\it Trigger}),
particle identification ({\it Baseline}), obvious background rejection
({\it Loose}), and optimized signal selection ({\it
Tight}). ``$Wc\bar{c}$'' refers to {\Wcc}+$Wcs$+{\Wss}
events. ``$Wjj$'' includes events with only light jets ($j$ = $u$,
$d$, $g$), and so all the tagged jets in this sample are either
mistagged or uninteresting ($\pi$ or $K$ decays). In the electron
channel, ``QCD'' refers to multijet events where a jet is mistaken for
an electron. In the muon channel, ``QCD'' refers to {\bbbar} events
where a nonisolated muon from a $b$~decay is mistaken for an isolated
muon. ``Other'' means multijet events with the isolated muon coming
from either a coincident cosmic ray, a mistake in pattern recognition,
or from a particle back-scattered off a beamline magnet into the
spectrometer.

\clearpage

It is clear from Table~\ref{table1} that, after trigger selection and
implementation of particle identification, most of the
electron-channel candidates correspond to multijet events with a jet
mistaken as an electron. In the muon channel, a similar fraction of
events have no true muon. After imposition of the tight criteria, the
next-largest background in the electron channel comes from $W$+jets
events with a fake muon tag.

Before other selections are applied to the single top quark signals,
the triggers are $40\%$ efficient in the electron channel and $60\%$
in the muon channel. In Run~2, it may be possible to lower thresholds
and include more objects in the trigger definitions, and thereby
improve the efficiencies without increasing rates from background.

Baseline selections require at least one object in the event to pass
the isolated electron or muon identification criteria, at least two
jets, and at least one good tagging muon. Only $28\%$ of electrons in
the original Monte Carlo signal samples pass the electron ID
requirements ($60\%$ after passing trigger thresholds). These criteria
are tight in order to minimize the rate from jets that mimic
electrons. Nevertheless, the sample is still dominated by multijet
events. We find that $30\%$ of the muons in the MC signal samples pass
the isolated muon ID requirements ($44\%$ after trigger thresholds are
passed). Contributions to these inefficiencies from geometrical
acceptance will not change much in Run~2, but lowering the transverse
momentum thresholds could significantly increase the efficiency.

The main difference between the loose selections in the two decay
channels is the rather severe one used to reject cosmic-ray
contamination in the muon channel. This has the additional effect of
rejecting much of the $W$+jets background, although it also throws out
$30\%$ of the single top quark events that fail no other
selection. If, despite the major detector improvements in Run~2, we
observe a significant fraction of events with false isolated-muons,
then more effort will have to go into either rejecting this background
offline, or measuring it and including it in the background
calculations.

To conclude, we have completed a search for single top quark
production at {\dzero}. The reader is referred to Ref.~1 for details
on the triggers, the three sets of selections, and for specifics on
particle identification. Limited statistics prevented the application
of tighter selections for rejecting background because of the need to
maximize the small signal acceptances. However, even under these
circumstances, we found it relatively straightforward to reduce the
{\ttbar} background, but, because of the large cross sections, it was
very difficult to reject $W$+jets events with false muon tags, and
events with false electrons or false isolated muons dominate the
background. These instrumental backgrounds may continue to limit
sensitivities to similar signals in Run~2.


\begin{thebibliography}{99.}

\bibitem{dzeropaper}
B.~Abbott {\it et al.}, {\dzero} Collaboration, 
submitted to Phys. Rev. Lett., hep-ex/0008024.

\bibitem{cdfpaper}
P.~Savard, to appear in the proceedings of the Meeting of the Division
of Particles and Fields of the American Physical Society, Columbus,
OH, August 2000.

\bibitem{willenbrock-s}
M.C.~Smith and S.~Willenbrock,
Phys. Rev. D {\bf 54}, 6696 (1996).

\bibitem{willenbrock-t}
T.~Stelzer, Z.~Sullivan, and S.~Willenbrock,
Phys. Rev. D {\bf 56}, 5919 (1997).

\bibitem{ttreview}
P.C.~Bhat, H.B.~Prosper, and S.S.~Snyder, Int. J. Mod. Phys. {\bf
A13}, 5113 (1998).

\bibitem{measVtb}
G.V.~Jikia and S.R.~Slabospitsky, Phys. Lett. B {\bf 295}, 136 (1992).

\end{thebibliography}
\end{document}